\documentclass[english,a4paper,prb,nofootinbib,nobibnotes]{revtex4}
\usepackage[T1]{fontenc}
\usepackage[latin9]{inputenc}
\usepackage{geometry}
\usepackage{babel}

\begin{document}

\title{Two-parameter complex Hadamard matrices for $N=6$ }

\author{Bengt R. Karlsson}

\email{bengt.karlsson@physics.uu.se}

\affiliation{Uppsala University, Dept of Physics and Astronomy, Box 516, SE-751
20 Uppsala, Sweden }

\begin{abstract}
A new, two-parameter, nonaffine family of complex Hadamard matrices
of order 6 is reported. It interpolates between the two Fourier families,
and contains as one-parameter subfamilies the Di{\c{t}}{\v{a}}
family, a symmetric family and an almost (up to equivalence) self-adjoint
family.
\end{abstract}

\maketitle

\section{introduction}

Complex Hadamard matrices have recently been given some attention
in connection with the search for mutually unbiased bases (MUBs).
In particular, it has been emphasized\citet{bengtsson} that the seach
for the maximum number of such bases in dimension 6 would be simplified
if a complete characterization of the related complex Hadamard matrices
were available (for notation, and a catalogue of complex Hadamard
matrices, see Refs. $2$ and $3$). There are indications\citet{bengtsson,Skinner}
that such a characterization will involve (at least) one four-parameter
family of Hadamard matrices, together with a single, isolated matrix
$S_{6}^{(0)}$. Until recently, however, the largest families known
were the two-parameter Fourier $F_{6}^{\,(2)}$ and Fourier transposed
$(F_{6}^{\,(2)})^{T}$ families. Three smaller, one-parameter, families
were also known, the $D_{6}^{(1)}$ family \citep{dita}, a symmetric
family\citep{Mat-Sz}, $M_{6}^{(1)}$, and a self-adjoint family\citep{beau_nic},
$B_{6}^{(1)}$, leaving the anticipated full set of complex Hadamards
of order 6 largely unexplored.

In this note, a new two-parameter family is constructed that interpolates
between the two Fourier families, and contains $D_{6}^{(1)}$ and
$M_{6}^{(1)}$ as subfamilies; futhermore, another newly found two-parameter
family\citet{Sz,Brier-Weig}, $X_{6}^{(2)}$ , can be seen as an extension
of $B_{6}^{(1)}$.%
\footnote{With the original parametrization\citep{Sz}, $(X_{6}^{\,(2)})^{T}$  is not
equivalent to $X_{6}^{\,(2)}$. However, topologically $(X_{6}^{\,(2)})^{T}$
and $X_{6}^{\,(2)}$ combine to form the surface of a sphere \citep{Brier-Weig}.
In this note, the parameter space of $X_{6}^{\,(2)}$ is understood
to have been extended to include also that of $(X_{6}^{\,(2)})^{T}$. %
} With these new results, the set of known complex Hadamard matrices
of order 6 has been significantly enlarged, and a more coherent family
pattern has emerged: there exist four (partially overlapping) two-parameter
families (the affine families $F_{6}^{\,(2)}$ and $(F_{6}^{\,(2)})^{T}$,
the nonaffine family reported here, and the nonaffine family $X_{6}^{(2)}$),
and all the previously known one-parameter families appear as subfamilies. 

On the other hand, by numerical means it is easy to generate matrices
which appear to belong to additional families, albeit of unknown parametric
form. Since these families also seem to have elements in common with
the known families, there is little doubt that eventually some (or
all) of the presently known two-parameter families will reappear as
sub-families of some yet to be found three- or four-parameter family
or families. 

The new family reported here, together with the family $X_{6}^{(2)}$,
are the only two-parameter, nonaffine families of complex Hadamard
matrices of any order $<12\,\,$ that have been found so far. For
order $12$, the four two-parameter families of order 6 can be combined
into nine-parameter families with up to four nonaffine parameters.

\section{Complex Hadamard matrices}

An $N\times N$ matrix $H$ with complex elements $h_{ij}$ is Hadamard
if all elements have modulus one, $|h_{ij}|=1$, and if $H^{\dagger}H/N=HH^{\dagger}/N=1$
(the unitarity constraint). Two Hadamard matrices are termed equivalent,
$H_{1}\sim H_{2}$, if they can be related through \begin{equation}
H_{2}=D_{2}P_{2}H_{1}P_{1}D_{1}\end{equation}
where $D_{1}$ and $D_{2}$ are diagonal unitary matrices, and $P_{1}$
and $P_{2}$ are permutation matrices. A set of equivalent Hadamard
matrices can be represented by a dephased matrix, with ones in the
first row and the first column. 

In the $6\times6$ case, several families of (non-equivalent) Hadamard
matrices are known. The two-parameter Fourier family $F_{6}^{\,(2)}$
can be given on the dephased form\citep{guide,webguide} \begin{equation}
F_{6}^{(2)}(a,b)=\left(\begin{array}{cccccc}
1 & \,\,\,1 & \,\,\quad\,\,\,1 & \,\,\,1 & \,\qquad\,1 & \,\,\,1\\
1 & z_{1}f & -z_{2}\bar{f} & -1 & -z_{1}f & z_{2}\bar{f}\\
1 & -\bar{f} & \,\,\,-f & \,\,\,1 & \,\,\,-\bar{f} & -f\\
1 & -z_{1} & \,\,\,\,\,\,\, z_{2} & -1 & \,\,\,\,\,\,\, z_{1} & -z_{2}\\
1 & -f & \,\,\,-\bar{f} & \,\,\,1 & \,\,\,\,-f & -\bar{f}\\
1 & z_{1}\bar{f} & -z_{2}f & -1 & -z_{1}\bar{f} & z_{2}f\end{array}\right)\label{eq:Fourier}\end{equation}
where $z_{1}=\exp(ia)$, $z_{2}=\exp(ib)$, $f=\exp({2\pi i/6})$,
and where $\bar{f}$ denotes the complex conjugate of $f$. The matrices
$(F_{6}^{(2)}(a,b))^{T}$, where $T$ denotes transposition, form
a separate family, with $(F_{6}^{(2)}(0,0))^{T}=F_{6}^{(2)}(0,0)\sim F_{6}$
as the generic Fourier matrix.

The one-parameter family $D_{6}^{(1)}$ can be given on the form\citep{guide,webguide}
\begin{equation}
D_{6}^{(1)}(c)=\left(\begin{array}{cccccc}
1 & \,\,\,1 & \,\,\,\,1 & \,\,\,\,1 & \,\,\,\,1 & \,\,\,\,1\\
1 & -1 & \,\,\,\, i & \,-i & \,-i & \,\,\,\, i\\
1 & \,\,\, i & \,-1 & \,\,\, iz & -iz & \,-i\\
1 & -i & \,\,\, i\bar{z} & \,-1 & \,\,\,\, i & -i\bar{z}\\
1 & -i & -i\bar{z} & \,\,\,\, i & \,-1 & \,\,\, i\bar{z}\\
1 & \,\,\, i & \,-i & -iz & \,\,\, iz & \,-1\end{array}\right)\label{eq:Dita}\end{equation}
where $z=\exp(ic)$ and $-\pi/4\le c\le\pi/4$.

\section{A new two-parameter family}

With the goal of constructing a new family of $6\times6$ complex
Hadamard matrices that interpolates between the two Fourier families,
consider the ansatz

\begin{equation}
H(x_{1},x_{2})=\left(\begin{array}{cccccc}
1 & 1 & 1 & 1 & 1 & 1\\
1 & -1 & z_{1} & -z_{1} & z_{1} & -z_{1}\\
1 & z_{2} & a_{11} & a_{12} & b_{11} & b_{12}\\
1 & -z_{2} & a_{21} & a_{22} & b_{21} & b_{22}\\
1 & z_{2} & c_{11} & c_{12} & d_{11} & d_{12}\\
1 & -z_{2} & c_{21} & c_{22} & d_{21} & d_{22}\end{array}\right)\label{eq:ansatz}\end{equation}
where all matrix elements are complex numbers of modulus one, and
where in particular $z_{1}=\exp(ix_{1})$ and $z_{2}=\exp(ix_{2})$.
 The matrices $H(x_{1},x_{2})$ will form a two-parameter Hadamard
family if, by imposing the unitarity constraint $H^{\dagger}H/6=HH^{\dagger}/6=1$,
all matrix elements can be solved for in terms of $z_{1}$ and $z_{2}$.

The conditions on the matrix elements $a_{ij}$, $b_{ij}$, $c_{ij}$
and $d_{ij}$ that follow from the unitarity constraint are of two
types, linear and quadratic. The linear conditions can be summarized as 
\begin{equation}
\mathbf{a}+\mathbf{c}=\mathbf{b}+\mathbf{d}=\mathbf{a}+\mathbf{b}=\mathbf{c}+\mathbf{d}=-\mathbf{Z}\label{eq:Linear_line}
\end{equation}
where $\mathbf{a}$, $\mathbf{b}$, $\mathbf{c}$ and $\mathbf{d}$
are $2\times2$ matrices with elements $a_{ij}$ etc, $i,j=1,2$, and where 
\begin{equation}
\mathbf{Z}=\left(\begin{array}{cc}
1-\frac{1}{2}(1-z_{1})(1-z_{2})\,\,\,\,\,\, & z_{2}(1-\frac{1}{2}(1-z_{1})(1-\bar{z}_{2}))\\
\\z_{1}(1-\frac{1}{2}(1-\bar{z}_{1})(1-z_{2}))\,\,\,\,\, & -z_{1}z_{2}(1-\frac{1}{2}(1-\bar{z}_{1})(1-\bar{z}_{2}))\end{array}\right)\label{eq:Z}
\end{equation}
The matrix $\mathbf{Z}$ satisfies the relations 
\begin{equation}
\mathbf{Z}^{\dagger}\mathbf{Z}=\mathbf{Z}\mathbf{Z}^{\dagger}=2\left(\begin{array}{cc}
1 & 0\\
0 & 1\end{array}\right)\label{eq:Z unitary_1}
\end{equation}
and its elements have the properties
\begin{eqnarray}
Z_{21} & = & z_{1}z_{2}\bar{Z}_{12}\nonumber \\
Z_{22} & = & -z_{1}z_{2}\bar{Z}_{11}
\end{eqnarray}
and 
\begin{eqnarray}
|Z_{11}|^{2} & = & |Z_{22}|^{2}=1-\frac{1}{4}(z_{1}-\bar{z}_{1})(z_{2}-\bar{z}_{2})\nonumber \\
|Z_{12}|^{2} & = & |Z_{21}|^{2}=1+\frac{1}{4}(z_{1}-\bar{z}_{1})(z_{2}-\bar{z}_{2})
\end{eqnarray}
From (\ref{eq:Linear_line}) it follows that $\mathbf{d}=\mathbf{a}$
and $\mathbf{c}=\mathbf{b}$, and it is therefore sufficient to proceed
with the simplified ansatz 
\begin{equation}
H(x_{1},x_{2})=\left(\begin{array}{cccccc}
1 & 1 & 1 & 1 & 1 & 1\\
1 & -1 & z_{1} & -z_{1} & z_{1} & -z_{1}\\
1 & z_{2} & a_{11} & a_{12} & b_{11} & b_{12}\\
1 & -z_{2} & a_{21} & a_{22} & b_{21} & b_{22}\\
1 & z_{2} & b_{11} & b_{12} & a_{11} & a_{12}\\
1 & -z_{2} & b_{21} & b_{22} & a_{21} & a_{22}\end{array}\right)\label{eq:ansatz2}
\end{equation}
for which the linear unitarity constraint reads
\begin{equation}
\mathbf{a}+\mathbf{b}=-\mathbf{Z}.\label{eq:Linear_short}
\end{equation}
The remaining, quadratic constraints can now be combined to read
\begin{equation}
(\mathbf{a^{\dagger}}+\mathbf{b^{\dagger}})(\mathbf{a}+\mathbf{b})=(\mathbf{a}+\mathbf{b})(\mathbf{a^{\dagger}}+\mathbf{b^{\dagger}})=2\left(\begin{array}{cc}
1 & 0\\
0 & 1\end{array}\right)\label{eq:non-lin_1}
\end{equation}
\begin{equation}
(\mathbf{a^{\dagger}}-\mathbf{b^{\dagger}})(\mathbf{a}-\mathbf{b})=(\mathbf{a}-\mathbf{b})(\mathbf{a^{\dagger}}-\mathbf{b^{\dagger}})=6\left(\begin{array}{cc}
1 & 0\\
0 & 1\end{array}\right)\label{eq:non-lin_2}
\end{equation}
In view of (\ref{eq:Z unitary_1}), the constraints (\ref{eq:non-lin_1})
are satified for any $\mathbf{a}$ and $\mathbf{b}$ satifying (\ref{eq:Linear_short}).

Since all elements of $\mathbf{a}$ and $\mathbf{b}$ are of modulus
one, and $|Z_{ij}|\le\sqrt{2}$ for all $i$ and $j$, the relation
(\ref{eq:Linear_short}) can be solved element by element,
\begin{eqnarray}
a_{ij} & = & -Z_{ij}(\frac{1}{2}+\sigma_{ij}\,\, i\sqrt{\frac{1}{|Z_{ij}|^{2}}-\frac{1}{4}})\nonumber \\
b_{ij} & = & -Z_{ij}(\frac{1}{2}-\sigma_{ij}\,\, i\sqrt{\frac{1}{|Z_{ij}|^{2}}-\frac{1}{4}})\label{eq:solution}
\end{eqnarray}
where the $\sigma_{ij}$'s are (so far undetermined) sign factors.
Furthermore, the remaining quadratic constraints (\ref{eq:non-lin_2})
are also satisfied by these solutions if only 
\begin{equation}
\sigma_{11}\sigma_{21}=\sigma_{12}\sigma_{22}.\label{eq:sigma_constraint}
\end{equation}
In all, therefore, there are $2^{4}/2=8$ sign combinations for the
$\sigma_{ij}'$s for which the ansatz (\ref{eq:ansatz}) leads to
a Hadamard matrix. For each choice of $x_{1}$ and $x_{2}$, the corresponding
$8$ matrices can be obtained one from the other through permutation
of rows and/or columns (3 and 5 and/or 4 and 6), and they are therefore
equivalent. The matrix with $\sigma_{11}=-\sigma_{22}=1$ and $\sigma_{12}=-\sigma_{21}=1$
is chosen as representative for the equivalence class (with this choice,
$\mathbf{a}$ and $\mathbf{b}$ become Hadamard matrices). \emph{The
end result is therefore a single, and new, two-parameter family of
complex Hadamard matrices of order $6$}.

\bigskip

In order to better expose the relationships between the elements of
$\mathbf{a}$ and $\mathbf{b}$, and between the new family and the
two Fourier families, introduce the notation
\begin{equation}
\begin{array}{rclcrcl}
f_{1} & = & -a_{11}(x_{1},x_{2}) & \,\,\,\,\,\,\,\,\,\,\,\,\,\,\,\, & f_{3} & = & -a_{11}(-x_{1},-x_{2})\\
f_{2} & = & -a_{11}(x_{1},-x_{2}) &  & f_{4} & = & -a_{11}(-x_{1},x_{2})\end{array}
\end{equation}
where, from (\ref{eq:solution}),
\begin{eqnarray}
a_{11}(x_{1},x_{2})=-(1-\frac{1}{2}(1-z_{1})(1-z_{2}))(\frac{1}{2}+i\sqrt{\frac{1}{1-\frac{1}{4}(z_{1}-\bar{z}_{1})(z_{2}-\bar{z}_{2})}-\frac{1}{4}})\nonumber \\
=-e^{i(x_{1}+x_{2})/2}(\cos(\frac{x_{1}-x_{2}}{2})-i\sin(\frac{x_{1}+x_{2}}{2}))(\frac{1}{2}+i\sqrt{\frac{1}{1+\sin(x_{1})\sin(x_{2})}-\frac{1}{4}})\nonumber \\
 &  & \mbox{}
\end{eqnarray}
In this notation,
\begin{equation}
\mathbf{a}=\left(\begin{array}{cc}
-f_{1}\, & -z_{2}f_{2}\\
-z_{1}\bar{f}_{2} & \,\,\, z_{1}z_{2}\bar{f}_{1}\end{array}\right)\,\,\,\,\,\,\,\,\,\,\mathrm{and}\,\,\,\,\,\,\,\,\,\mathbf{b}=\left(\begin{array}{cc}
-\bar{f}_{3}\, & -z_{2}\bar{f}_{4}\\
-z_{1}f_{4} & \,\,\, z_{1}z_{2}f_{3}\end{array}\right)
\end{equation}
The factors $f_{i}$ are of unit modulus, and in the limits $z_{1}\to1$
and/or $z_{2}\to1$, i.e. $x_{1}\to0$ and/or $x_{2}\to0$, they all
reduce to the factor $f=(1+i\sqrt{3})/2=\exp(2\pi i/6)$ appearing
in (\ref{eq:Fourier}).

\bigskip

It should finally be noted that $a_{11}(x_{1}+\pi,\, x_{2})=z_{2}a_{11}(x_{1,}-x_{2})$
and $a_{11}(x_{1},\, x_{2}+\pi)=z_{1}a_{11}(-x_{1},\, x_{2})$. As
a result, $H(x_{1}+\pi,\, x_{2})=H(x_{1},x_{2})P_{34}P_{56}$ and
$H(x_{1},x_{2}+\pi)=P_{36}P_{45}H(x_{1},x_{2})$ where $P_{ij}$ is
the $i\leftrightarrow j$ (row or column) permutation matrix, i.e.
$H(x_{1}+\pi,x_{2})$ and $H(x_{1},x_{2}+\pi)$ are both equivalent
to $H(x_{1},x_{2})$. Hence, for the new family it is sufficient to
chose the parameters from the domain $-\frac{\pi}{2}<x_{1}\le\frac{\pi}{2}$,
$-\frac{\pi}{2}<x_{2}\le\frac{\pi}{2}$, where as before $z_{1}=\exp(ix_{1})$
and $z_{2}=\exp(ix_{2})$.

\section{One-parameter subfamilies}

Some one-parameter subfamilies of the new two-parameter family are
of particular interest. 

\subsubsection*{1. Two Fourier subfamilies.}

Taking $x_{1}=0$ or $x_{2}=0$ in $H(x_{1},x_{2})$ one finds 
\begin{eqnarray}
H(x,0) & \sim & F_{6}^{(2)}(x,\, x)\nonumber \\
H(0,x) & \sim & (F_{6}^{(2)}(x,\, x))^{T}
\end{eqnarray}
where $F_{6}^{(2)}(x,x)$ is a subfamily of the Fourier family $F_{6}^{\,(2)}(a,b)$.
For instance, taking $x_{2}=0$, it follows that $f_{1}=f_{2}=f_{3}=f_{4}=f$,
so that 
\begin{equation}
\mathbf{a}=\left(\begin{array}{cc}
\,\,\,-f & \,-f\\
-z_{1}\bar{f}\, & \, z_{1}\bar{f}\end{array}\right)\,\,\,\,\,\,\,\mathrm{and}\,\,\,\,\,\,\,\mathbf{b}=\left(\begin{array}{cc}
\,\,\,-\bar{f} & \,\,-\bar{f}\\
-z_{1}f & \, z_{1}f\end{array}\right)
\end{equation}
and the one-parameter subfamily $F_{6}^{(2)}(x,x)$ is obtained as
$P_{26}P_{24}P_{35}H(x,0)P_{24}P_{35}$. 

The new family therefore interpolates between (subsets of) the Fourier
families, as intended, with for instance $H_{I}(\xi)=H(\xi x,(1-\xi)x),\,\,\,0\le\xi\le1$,
as an interpolating subfamily for any given $x$.

\subsubsection*{2. A symmetric subfamily.}

A symmetric subfamily $H(x,x)$ is found along the main diagonal of
the parameter domain. In this case $z_{1}=z_{2}=z=\exp(ix)$ so that
\begin{eqnarray}
f_{1} & \to & z(1-i\sin(x))(\frac{1}{2}+i\sqrt{\frac{1}{1+\sin^{2}(x)}-\frac{1}{4}})\equiv zg_{1}\nonumber \\
f_{2} & = & f_{4}\to\cos(x)(\frac{1}{2}+i\sqrt{\frac{1}{\cos^{2}(x)}-\frac{1}{4}})\equiv g_{2}\label{eq:g_def}\\
f_{3} & \to & \bar{z}(1+i\sin(x))(\frac{1}{2}+i\sqrt{\frac{1}{1+\sin^{2}(x)}-\frac{1}{4}})\equiv\bar{z}g_{3}\nonumber \end{eqnarray}
and
\begin{equation}
\mathbf{a}=z\left(\begin{array}{cc}
\,-g_{1} & \,-g_{2}\\
-\bar{g}_{2} & \,\,\,\,\bar{g}_{1}\end{array}\right)\,\,\,\,\,\,\,\mathrm{and}\,\,\,\,\,\,\,\mathbf{b}=z\left(\begin{array}{cc}
-\bar{g}_{3} & -\bar{g}_{2}\\
-g_{2} & \,\,\, g_{3}\end{array}\right).
\end{equation}
The matrix obtained after permutation of rows 4 and 6 is symmetric,
\begin{equation}
P_{46}H(x,x)=\left(\begin{array}{cccccc}
1 & \,\,\,1 & \,\,\,1 & \,\,\,1 & \,1 & \,\,\,1\\
1 & -1 & \,\,\, z & -z & \: z & -z\\
1 & \,\,\, z & \,-zg_{1} & \,\,-zg_{2} & -z\bar{g}_{3} & \,\,-z\bar{g}_{2}\\
1 & -z & \,-zg_{2} & \,\,\,\,\, zg_{3} & -z\bar{g}_{2} & \,\,\,\,\, z\bar{g}_{1}\\
1 & \,\,\, z & \,-z\bar{g}_{3} & \,\,-z\bar{g}_{2} & -zg_{1} & \,\,-zg_{2}\\
1 & -z & \,-z\bar{g}_{2} & \,\,\,\,\,\, z\bar{g}_{1} & -zg_{2} & \,\,\,\,\, zg_{3}\end{array}\right).
\end{equation}
and this subfamily coincides with the symmetric family $M_{6}^{(1)}$
recently reported by Matolcsi and Szöll\H osi\citep{Mat-Sz}. Specifically,
$P_{46}H(x,x)$ equals $M_{6}^{(1)}(x)$ after permutation of rows
4 and 5, and of columns 4 and 5.

\subsubsection*{3. An essentially self-adjoint subfamily}

Another subfamily of interest is $H(x,-x)$, found along one of the
diagonals in the parameter domain. In this case $z_{2}=\bar{z}_{1}$,
so that (with $z_{1}=z=\exp(ix)$)
\begin{eqnarray}
f_{1} & = & f_{3}\to\cos(x)(\frac{1}{2}+i\sqrt{\frac{1}{\cos^{2}(x)}-\frac{1}{4}})=g_{2}\nonumber \\
f_{2} & \to & z(1-i\sin(x))(\frac{1}{2}+i\sqrt{\frac{1}{1+\sin^{2}(x)}-\frac{1}{4}})=zg_{1}\nonumber \\
f_{4} & \to & \bar{z}(1+i\sin(x))(\frac{1}{2}+i\sqrt{\frac{1}{1+\sin^{2}(x)}-\frac{1}{4}})=\bar{z}g_{3}
\end{eqnarray}
and
\begin{equation}
\mathbf{a}=\left(\begin{array}{cc}
-g_{2} & -g_{1}\\
-\bar{g}_{1} & \,\,\,\bar{g}_{2}\end{array}\right)\,\,\,\,\,\,\,\mathrm{and}\,\,\,\,\,\,\,\mathbf{b}=\left(\begin{array}{cc}
-\bar{g}_{2} & -\bar{g}_{3}\\
-g_{3} & \,\,\, g_{2}\end{array}\right).
\end{equation}
where $g_{1}$, $g_{2}$ and $g_{3}$ are the factors defined in (\ref{eq:g_def}).
The resulting one-parameter Hadamard matrix is not self-adjoint, but
essentially self-adjoint in the sense that it is equivalent to its
adjoint, $[H(x,-x)]^{\dagger}\sim H(x,-x)$. Indeed, $[H(x,-x)]^{\dagger}$
only differs from $H(x,-x)$ through an interchange of columns 4 and
6, and of rows 3 and 5, 
\begin{equation}
H(x,-x)=\left(\begin{array}{cccccc}
1 & \,\,\,1 & 1 & \,\,\,1 & 1 & \,\,\,1\\
1 & -1 & z & -z & z & -z\\
1 & \,\,\,\bar{z} & -g_{2} & -g_{1} & -\bar{g}_{2} & -\bar{g}_{3}\\
1 & -\bar{z} & -\bar{g}_{1} & \,\,\,\bar{g}_{2} & -g_{3} & \,\,\, g_{2}\\
1 & \,\,\,\bar{z} & -\bar{g}_{2} & -\bar{g}_{3} & -g_{2} & -g_{1}\\
1 & -\bar{z} & -g_{3} & \,\,\, g_{2} & -\bar{g}_{1} & \,\,\,\bar{g}_{2}\end{array}\right)=P_{35}[H(x,-x)]^{\dagger}P_{46}.
\end{equation}

\subsubsection*{4. The Di{\c{t}}{\v{a}} family $D_{6}^{(1)}$.}

At the corners of the parameter domain, the new family contains $D_{6}^{(1)}$
as a one-parameter subfamily. Indeed, taking $x_{1}=\pi/2-\epsilon_{1}$
and $x_{2}=\pi/2-\epsilon_{2}$ , and letting $\epsilon_{1},\epsilon_{2}\to0^{+}$,
one finds
\begin{eqnarray}
f_{1} & \to & i\nonumber \\
f_{2} & \to & i\frac{\epsilon_{1}+\epsilon_{2}+i(\epsilon_{1}-\epsilon_{2})}{\sqrt{2(\epsilon_{1}^{2}+\epsilon_{2}^{2})}}\to iz\nonumber \\
f_{3} & \to & 1\nonumber \\
f_{4} & \to & i\frac{\epsilon_{1}+\epsilon_{2}-i(\epsilon_{1}-\epsilon_{2})}{\sqrt{2(\epsilon_{1}^{2}+\epsilon_{2}^{2})}}\to i\bar{z}
\end{eqnarray}
where $z=\exp(ix)$ and
\begin{equation}
x=\lim_{\epsilon_{1},\epsilon_{2}\to0^{+}}\arctan(\frac{\epsilon_{1}-\epsilon_{2}}{\epsilon_{1}+\epsilon_{2}}).
\end{equation}
The angle $x$ can take any value between $-\pi/4$ and $\pi/4$ depending
on how the  limit point is approached. The resulting one-parameter
Hadamard family 
\begin{equation}
\left(\begin{array}{cccccc}
1 & \,\,\,1 & \,\,\,1 & \,\,\,1 & \,\,\,1 & \,\,\,1\\
1 & -1 & \,\,\, i & -i & \,\,\, i & -i\\
1 & \,\,\, i & -i & \,\,\, z & -1 & -z\\
1 & -i & -\bar{z} & \,\,\, i & \,\,\,\bar{z} & -1\\
1 & \,\,\, i & -1 & -z & -i & \,\,\, z\\
1 & -i & \,\,\,\bar{z} & -1 & -\bar{z} & \,\,\, i\end{array}\right)
\end{equation}
is equivalent to the family $D_{6}^{(1)}(c)$ in (\ref{eq:Dita}).
To see this, first multiply the $6$ columns by $1,-1,-i,i,-i$ and
$i$, respectively, and then interchange rows $1$ and 2, rows 3 and
4, and columns 3 and 4. If finally $x$ is replaced by $-c$, the
$D_{6}^{(1)}(c)$ of (\ref{eq:Dita}) is obtained.

\subsubsection*{5. Other subfamilies.}

Other particularly simple subfamilies are found along the borders
of the parameter domain, $H(x,\frac{\pi}{2})$ and $H(\frac{\pi}{2},x)$.

\section{Summary and outlook}

Collecting results, a new two-parameter family of complex Hadamard
matrices of order 6 has been found,
\begin{equation}
H(x_{1},x_{2})=\left(\begin{array}{cccccc}
1 & \,\,\,1\,\, & 1 & 1 & 1 & 1\\
1 & -1\,\, & z_{1} & -z_{1} & z_{1} & -z_{1}\\
1 & \,\,\, z_{2}\,\, & -f_{1}\, & -z_{2}f_{2} & -\bar{f}_{3}\, & -z_{2}\bar{f}_{4}\\
1 & -z_{2}\,\, & -z_{1}\bar{f}_{2} & \,\,\, z_{1}z_{2}\bar{f}_{1} & -z_{1}f_{4} & \,\,\, z_{1}z_{2}f_{3}\\
1 & \,\,\, z_{2}\,\, & -\bar{f}_{3}\, & -z_{2}\bar{f}_{4} & -f_{1}\, & -z_{2}f_{2}\\
1 & -z_{2}\,\, & -z_{1}f_{4} & \,\,\, z_{1}z_{2}f_{3} & -z_{1}\bar{f}_{2} & \,\,\, z_{1}z_{2}\bar{f}_{1}\end{array}\right)
\end{equation}
with parameters $z_{1}=\exp(ix_{1})$ and $z_{2}=\exp(ix_{2})$, $-\frac{\pi}{2}<x_{1}\le\frac{\pi}{2}$
,$-\frac{\pi}{2}<x_{2}\le\frac{\pi}{2}$. The elements are given in
terms of four factors,
\begin{equation}
\begin{array}{rclcrcl}
f_{1} & = & f(x_{1},x_{2}) & \,\,\,\,\,\,\,\,\,\,\,\,\,\,\,\, & f_{3} & = & f(-x_{1},-x_{2})\\
f_{2} & = & f(x_{1},-x_{2}) &  & f_{4} & = & f(-x_{1},x_{2})\end{array}
\end{equation}
where
\begin{eqnarray}
f(x_{1},x_{2})=(1-\frac{1}{2}(1-z_{1})(1-z_{2}))(\frac{1}{2}+i\sqrt{\frac{1}{1-\frac{1}{4}(z_{1}-\bar{z}_{1})(z_{2}-\bar{z}_{2})}-\frac{1}{4}})\nonumber \\
=e^{i(x_{1}+x_{2})/2}(\cos(\frac{x_{1}-x_{2}}{2})-i\sin(\frac{x_{1}+x_{2}}{2}))(\frac{1}{2}+i\sqrt{\frac{1}{1+\sin(x_{1})\sin(x_{2})}-\frac{1}{4}})\nonumber \\
 &  & \mbox{}
\end{eqnarray}
with $|f(x_{1},x_{2})|=1$.

The new family has the subfamily $H(x,0)$ in common with the Fourier
family, the subfamily $H(0,x)$ in common with the Fourier-transposed
family, a symmetric subfamily $H(x,x)$ that coincides with $M_{6}^{(1)}$,
and it contains the Di{\c{t}}{\v{a}} family $D_{6}^{(1)}$ at
the points $x_{1}=\pm\frac{\pi}{2},\,\,\, x_{2}=\pm\frac{\pi}{2}$.
It therefore provides a bridge between these previously studied families.

Other simple subfamilies include the border families $H(x,\frac{\pi}{2})$
and $H(\frac{\pi}{2},x)$. The subfamily $H(x,-x)$ is essentially
self-adjoint in the sense that each member matrix is equivalent to
its adjoint, $[H(x,-x)]^{\dagger}\sim H(x,-x)$. 

Using well-known constructions, the four two-parameter families $F_{6}^{\,(2)}$,
$(F_{6}^{\,(2)})^{T}$, $H$ and $X_{6}^{(2)}$  may be combined into
multi-parameter complex Hadamard matrices of higher orders. For instance,
let $H_{1}(x_{1},x_{2})$ and $H_{2}(x_{3},x_{4})$ be chosen among
the four families of order $6$, and let $D=\mathrm{diag}(1,e^{i\delta_{1}},...,e^{i\delta_{5}})$.
Then the matrices 
\begin{equation}
\left(\begin{array}{cc}
H_{1}(x_{1},x_{2}) & DH_{2}(x_{3},x_{4})\\
H_{1}(x_{1},x_{2}) & -DH_{2}(x_{3},x_{4})\end{array}\right)
\end{equation}
form nine-parameter families of (dephased) complex Hadamard matrices,
significantly extending the list\citep{webguide} of order 12 matrices.


\begin{thebibliography}{9}

\bibitem[1]{bengtsson} I. Bengtsson, W. Bruzda, Å.~Ericsson, J-Å.
Larsson, W.~Tadej, and K.~{\.{Z}}yczkowski, \textsl{\emph{J. Math.
Phys.}} \textbf{\textsl{\emph{48}}}\textsl{\emph{, 052106 (2007).}}

\bibitem[2]{guide} W. Tadej and K. {\.{Z}}yczkowski, Open Syst.
\& Inf. Dyn. \textbf{13}, 133 (2006); e-print arXiv:quant-ph/0512154.

\bibitem[3]{webguide} W. Tadej and K. {\.{Z}}yczkowski, http://chaos.if.uj.edu.pl/$\sim$karol/hadamard.

\bibitem[4]{Skinner} A. J. Skinner, V. A. Newell and R. Sanchez,
J. Math. Phys. \textbf{\textsl{\emph{50}}}\textsl{\emph{, 012107 (2009).}}

\bibitem[5]{dita} P. Di{\c{t}}{\v{a}}, J. Phys. A \textbf{37},
5355 (2004).

\bibitem[6]{Mat-Sz} M. Matolcsi and F. Szöll\H osi, Open Syst. \&
Inf. Dyn. \textbf{\textsl{\emph{15:2}}}, 93 (2008); e-print arXiv:math/0702043v1.

\bibitem[7]{beau_nic} K. Beauchamp and R. Nicoara, Linear Algebra
Appl. \textbf{428}, No. 8-9, 1833 (2008).

\bibitem[8]{Sz} F. Szöll\H osi, e-print arXiv:math/08113930v1. 

\bibitem[9]{Brier-Weig} S. Brierley and S. Weigert, e-print arXiv:quant-ph/0901.4051v1.

\end{thebibliography}
\end{document}